\documentclass[english,10pt,twocolumn,twoside]{IEEEtran}
\usepackage[T1]{fontenc}
\usepackage[latin9]{inputenc}
\usepackage{bm}
\usepackage{amsmath}
\usepackage{amssymb}
\usepackage{graphicx}

\makeatletter
\usepackage{colortbl}
 \usepackage{cite}

\makeatother

\usepackage{babel}
\begin{document}

\title{Non-colocated Time-Reversal MUSIC:\linebreak{}
 High-SNR Distribution of Null Spectrum }

\author{D.~Ciuonzo,~\IEEEmembership{Senior~Member,~IEEE} and P. Salvo~Rossi,~\IEEEmembership{Senior~Member,~IEEE}\thanks{Manuscript received 2nd December 2016; accepted 25th January 2017.\protect \\
D. Ciuonzo is with DIETI, University of Naples ``Federico II'',
Naples, Italy. \protect \\
P. Salvo Rossi is with the Dept. of Electronics and Telecommunications,
NTNU, Trondheim, Norway. \protect \\
E-mail: \{domenico.ciuonzo, salvorossi\}@ieee.org.}\vspace{-0.8cm}
}
\maketitle
\begin{abstract}
We derive the asymptotic distribution of the null spectrum of the
well-known Multiple Signal Classification (MUSIC) in its computational
Time-Reversal (TR) form. The result pertains to a single-frequency
non-colocated multistatic scenario and several TR-MUSIC variants are
here investigated. The analysis builds upon the 1st-order perturbation
of the singular value decomposition and allows a simple characterization
of null-spectrum moments (up to the 2nd order). This enables a comparison
in terms of spectrums stability. Finally, a numerical analysis is
provided to confirm the theoretical findings\footnotetext{\emph{Notation} - Lower-case (resp. Upper-case) bold letters denote column vectors (resp. matrices), with $a_{n}$ (resp. $a_{n,m}$) being the $n$th (resp. the $(n,m)$th) element of $\bm{a}$ (resp. $\bm{A}$); $\mathbb{E}\{\cdot\}$, $\mathrm{var}\{\cdot\}$, $(\cdot)^{T}$, $(\cdot)^{\dagger}$, $(\cdot)^{*}$, $\mathrm{Tr}\left[\cdot\right]$, $\mathrm{vec}(\cdot)$, $(\cdot)^{-}$, $\Re\left(\cdot\right)$, $\delta(\cdot)$, $\left\Vert \cdot\right\Vert _{F}$ and $\left\Vert \cdot\right\Vert $ denote expectation, variance, transpose, Hermitian, conjugate, matrix trace, vectorization, pseudo-inverse, real part, Kronecker delta, Frobenius and $\ell_{2}$ norm operators, respectively; $j$ denotes the imaginary unit; $\bm{0}_{N\times M}$ (resp. $\bm{I}_{N}$) denotes the $N\times M$ null (resp. identity) matrix; $\bm{0}_{N}$ (resp. $\bm{1}_{N}$) denotes the null (resp. ones) column vector of length $N$; $\mathrm{diag}(\bm{a})$ denotes the diagonal matrix obtained from the vector $\bm{a}$; $\bm{x}_{1:M}\triangleq\begin{bmatrix}\bm{x}_{1}^{T} & \cdots & \bm{x}_{M}^{T}\end{bmatrix}^{T}$ denotes the vector concatenation; $\mathcal{N}_{\mathbb{C}}(\bm{\mu},\bm{\Sigma})$  denotes a proper  complex Gaussian pdf with mean vector $\bm{\mu}$ and covariance $\bm{\Sigma}$; $\mathcal{C\chi}_{N}^{2}$ denotes a complex chi-square distribution with $N$ (complex) Degrees of Freedom (DOFs); finally the symbol $\sim$ means \textquotedblleft distributed as\textquotedblright.}.
\end{abstract}


\section{Introduction\label{sec: Introduction}}

\IEEEPARstart{T}{ime-Reversal} (TR) refers to all those methods which
exploit the invariance of the wave equation (in lossless and stationary
media) by re-transmitting a time-reversed version of the scattered
(or radiated) field measured by an array to focus on a scattering
object (or radiating source), by physical \cite{Fink1993} or synthetic
\cite{Cassereau1992} means. In the latter case (\emph{computational}
\emph{TR}), it consists in \emph{numerically} back-propagating the
field data by using a known Green\textquoteright s function, representative
of the propagation medium. Since the employed Green function depends
on the object (or source) position, an image is formed by varying
the probed location (this procedure is referred to as ``imaging'').
Computational TR has been successfully applied in different contexts
such as subsurface prospecting \cite{Micolau2003}, through-the-wall
\cite{Li2010} and microwave imaging \cite{Hossain2013}.

The key entity in TR-imaging is the Multistatic Data Matrix (MDM),
whose entries are the scattered field due to each Transmit-Receive
(Tx-Rx) pair. Two popular methods for TR-imaging are the decomposition
of TR operator (DORT) \cite{Prada1996} and the TR Multiple Signal
Classification (TR-MUSIC) \cite{Devaney2005}. DORT imaging exploits
the MDM spectrum by back-propagating each eigenvector of the so-called
\emph{signal subspace}, thus allowing to selectively focus on each
(well-resolved) scatterer. On the other hand, TR-MUSIC imaging is
based on a complementary point of view and relies on the\emph{ noise
subspace} (viz. orthogonal-subspace\footnote{Such term underlines that it is orthogonal to the signal subspace.}),
leading to satisfactory performance as long as the data space dimension
exceeds the signal subspace dimension and sufficiently high Signal-to-Noise-Ratio
(SNR) is present. TR-MUSIC was first introduced for a Born Approximated
(BA) linear scattering model \cite{Devaney2005} and, later, successfully
applied to the Foldy-Lax (FL) non-linear model \cite{Marengo2007}.
Also, it became popular mainly due to: ($a$) algorithmic efficiency;
($b$) no need for approximate scattering models; and ($c$) finer
resolution than the diffraction limits (especially in scenarios with
few scatterers). Recently, TR-MUSIC has been expanded to extended
scatterers in \cite{Marengo2007a}.

Though a vast literature on performance analysis of MUSIC \cite{Schmidt1986}
for Direction-Of-Arrival (DOA) estimation exists (see \cite{Kaveh1986,Friedlander1990}
for resolution studies and \cite{Porat1988,Stoica1989,Li1990,Swindlehurst1992}
for asymptotic Mean Squared Error (MSE) derivation, with more advanced
studies presented in \cite{Ferreol2006,Ferreol2008,Ferreol2010}),
such results cannot be directly applied to TR-MUSIC. Indeed, in TR
framework scatterers/sources are generally assumed deterministic and
more importantly a \emph{single snapshot is used}, whereas MUSIC results
for DOA refer to a different asymptotic condition (i.e. a large number
of snapshots). Also, to our knowledge, \emph{no corresponding theoretical
results have been proposed in the literature} for TR-MUSIC, except
for \cite{Ciuonzo2014,Ciuonzo2015}, providing the asymptotic (high-SNR)
localization MSE for point-like scatterers. Yet, a few works have
tackled achievable theoretical performance both for\emph{ }BA and
FL models via the Cramér-Rao lower-bound \cite{Shi2007a}.

In this letter we provide a null-spectrum\footnote{We underline that the MUSIC imaging function is commonly referred
to as ``pseudo-spectrum'' in DOA literature. Though less used, in
this paper we will instead adopt to the term ``null-spectrum'' employed
in \cite{Choi1993}, as the latter work represents the closest counterpart
in DOA estimation to the present study.} analysis of TR-MUSIC for point-like scatterers, via a 1st-order perturbation
of Singular Value Decomposition (SVD) \cite{Li1993}, thus having
asymptotic validity (i.e. meaning a high SNR regime). The present
results are based on a homogeneous background assumption and neglecting
mutual coupling, as well as polarization or antenna pattern effects.
Here we build upon \cite{CiuonzoSubmitted2016} (tackling the simpler
colocated case) and consider a \emph{general} non-colocated multistatic
setup with BA/FL models where several TR-MUSIC variants, proposed
in the literature, are here investigated. The obtained results complement
those found in DOA literature \cite{Choi1993} and allow to obtain
both the mean and the variance of each null-spectrum, as well as to
draw-out its pdf. Also, they highlight performance dependence of null-spectrum
on the scatterers/arrays configurations and compare TR-MUSIC variants
in terms of spectrum stability. We recall that stability property
is important for TR-MUSIC, and has been investigated by numerical
means \cite{Asgedom2011,Yavuz2008} or using compressed-sensing based
approaches \cite{Fannjiang2011}. Finally, a few numerical examples,
for a 2-D geometry with scalar scattering, are presented to confirm
our findings.

The letter is organized as follows: Sec.~\ref{sec:System-Model}
describes the system model and reviews classic results on SVD perturbation
analysis. Sec.~\ref{sec: Performance analysis} presents the theoretical
characterization of TR-MUSIC null-spectrum, whereas its validation
is shown in Sec.~\ref{sec: Numerical Results} via simulations. Finally,
conclusions are in Sec.~\ref{sec: Conclusions}.

\section{System Model\label{sec:System-Model}}

We consider localization of $M$ point-like scatterers\footnote{The number of scatterers $M$ is assumed to be known, as usually done
in array-processing literature \cite{Krim1996}.} at unknown positions $\{\bm{x}_{k}\}_{k=1}^{M}$ in $\mathbb{R}^{p}$
with unknown scattering potentials $\{\tau_{k}\}_{k=1}^{M}$ in $\mathbb{C}$.
The Tx (resp. Rx) array consists of $N_{T}$ (resp. $N_{R}$) isotropic
point elements (resp. receivers) located at $\{\tilde{\bm{r}}_{i}\}_{i=1}^{N_{T}}$
in $\mathbb{R}^{p}$ (resp. $\{\bar{\bm{r}}_{j}\}_{j=1}^{N_{R}}$
in $\mathbb{R}^{p}$). The illuminators first send signals to the
probed scenario (in a known homogeneous background with wavenumber
$\kappa$) and the transducer array records the received signals.
The (single-frequency) measurement model is then \cite{Shi2005}:
\begin{eqnarray}
\bm{K}_{n} & = & \bm{K}(\bm{x}_{1:M},\bm{\tau})+\bm{W}\\
 & = & \bm{G}_{\mathrm{r}}(\bm{x}_{1:M})\,\bm{M}(\bm{x}_{1:M},\bm{\tau})\,\bm{G}_{\mathrm{t}}(\bm{x}_{1:M})^{T}+\bm{W}\label{eq: Multistatic model explicit}
\end{eqnarray}
where $\bm{K}_{n}\in\mathbb{C}^{N_{R}\times N_{T}}$ (resp. $\bm{K}(\bm{x}_{1:M},\bm{\tau})$)
denotes the measured (resp. noise-free) MDM. Differently $\bm{W}\in\mathbb{C}^{N_{R}\times N_{T}}$
is a noise matrix s.t. $\mathrm{vec}(\bm{W})\sim\mathcal{N}_{\mathbb{C}}(\bm{0}_{N},\sigma_{w}^{2}\,\bm{I}_{N})$,
where $N\triangleq N_{T}N_{R}$. Additionally, we have denoted: ($i$)
the vector of scattering coefficients as $\bm{\tau}\triangleq\left[\begin{array}{ccc}
\tau_{1} & \cdots & \tau_{M}\end{array}\right]^{T}\in\mathbb{C}^{M\times1}$; ($ii$) ($b$) the Tx (resp. Rx) array matrix as $\bm{G}_{\mathrm{t}}(\bm{x}_{1:M})\in\mathbb{C}^{N_{T}\times M}$
(resp. $\bm{G}_{\mathrm{r}}(\bm{x}_{1:M})\in\mathbb{C}^{N_{R}\times M}$),
whose $(i,j)$th entry equals $\mathcal{G}(\tilde{\bm{r}}_{i},\bm{x}_{j})$
(resp. $\mathcal{G}(\bar{\bm{r}}_{i},\bm{x}_{j})$), where $\mathcal{G}(\cdot,\cdot)$
denotes the (scalar) background \emph{Green function }\cite{Devaney2005}.
Also, $j$th column $\bm{g}_{\mathrm{t}}(\bm{x}_{j})$ (resp. $\bm{g}_{\mathrm{r}}(\bm{x}_{j})$)
of $\bm{G}_{\mathrm{t}}(\bm{x}_{1:M})$ (resp. $\bm{G}_{\mathrm{r}}(\bm{x}_{1:M})$)
denotes the Tx (resp. Rx) Green's function vector evaluated at $\bm{x}_{j}$.
In Eq. (\ref{eq: Multistatic model explicit}) the scattering matrix
$\bm{M}(\bm{x}_{1:M},\bm{\tau})\in\mathbb{C}^{M\times M}$ equals
$\bm{M}(\bm{x}_{1:M},\bm{\tau})\triangleq\mathrm{diag}(\bm{\tau})$
for BA model \cite{Devaney2005}, while $\bm{M}(\bm{x}_{1:M},\bm{\tau})\triangleq\left[\mathrm{diag}^{-1}(\bm{\tau})-\bm{S}(\bm{x}_{1:M})\right]^{-1}$
in the case of FL model \cite{Shi2007a}, where the $(m,n)$th entry
of $\bm{S}(\bm{x}_{1:M})$ equals $\mathcal{G}(\bm{x}_{m},\bm{x}_{n})$
when $m\neq n$ and zero otherwise. We recall that our null-spectrum
analysis of TR-MUSIC \emph{is general} and can be applied to both
scattering models.

Finally, we define the $\mathrm{SNR\triangleq}\left\Vert \bm{K}(\bm{x}_{1:M},\bm{\tau})\right\Vert _{F}^{2}/(\sigma_{w}^{2}\,N_{T}N_{R})$
and, for notational convenience, $N_{\mathrm{Rdof}}\triangleq(N_{R}-M)$
and $N_{\mathrm{Tdof}}\triangleq(N_{T}-M)$ as the dimensions of the
left and right orthogonal subspaces, whereas $N_{\mathrm{dof}}\triangleq(N_{\mathrm{Rdof}}+N_{\mathrm{Tdof}})$.

\subsection{TR-MUSIC Spatial Spectrum}

Several TR-MUSIC variants have been proposed in the literature for
the non co-located setup \cite{Marengo2007}. A first approach consists
in using the so-called \emph{Rx mode TR-MUSIC}, which evaluates the
\emph{null (or spatial) spectrum} (assuming $M<N_{R}$):
\begin{equation}
\mathcal{P}_{\mathrm{r}}(\bm{x};\widetilde{\bm{U}}_{n})\triangleq\bar{\bm{g}}_{r}(\bm{x})^{\dagger}\,\widetilde{\bm{P}}_{\mathrm{r},n}\,\bar{\bm{g}}_{r}(\bm{x})=\left\Vert \widetilde{\bm{U}}_{n}^{\dagger}\,\bar{\bm{g}}_{r}(\bm{x})\right\Vert ^{2}\,,\label{eq: TR-MUSIC Rx Mode}
\end{equation}
where $\widetilde{\bm{U}}_{n}\in\mathbb{C}^{N_{R}\times N_{\mathrm{Rdof}}}$
is the matrix of left singular vectors of $\bm{K}_{n}$ spanning the
noise subspace, $\bar{\bm{g}}_{\mathrm{r}}(\bm{x})\triangleq\bm{g}_{\mathrm{r}}(\bm{x})/\left\Vert \bm{g}_{\mathrm{r}}(\bm{x})\right\Vert $
is the unit-norm Rx Green vector function and $\widetilde{\bm{P}}_{\mathrm{r},n}\triangleq(\widetilde{\bm{U}}_{n}\widetilde{\bm{U}}_{n}^{\dagger})$
(i.e. the ``noisy'' projector into the left noise subspace). A dual
approach, denoted as \emph{Tx mode TR-MUSIC}, constructs the null
spectrum (assuming $M<N_{T}$):
\begin{equation}
\mathcal{P}_{\mathrm{t}}(\bm{x};\widetilde{\bm{V}}_{n})\triangleq\bar{\bm{g}}_{\mathrm{t}}(\bm{x})^{T}\,\widetilde{\bm{P}}_{\mathrm{t},n}\,\bar{\bm{g}}_{\mathrm{t}}(\bm{x})^{*}=\left\Vert \widetilde{\bm{V}}_{n}^{\dagger}\,\bar{\bm{g}}_{\mathrm{t}}^{*}(\bm{x})\right\Vert ^{2}\,,\label{eq: TR-MUSIC Tx Mode}
\end{equation}
where $\widetilde{\bm{V}}_{n}\in\mathbb{C}^{N_{T}\times N_{\mathrm{Tdof}}}$
is the matrix of right singular vectors of $\bm{K}_{n}$ spanning
the noise subspace, $\bar{\bm{g}}_{\mathrm{t}}(\bm{x})\triangleq\bm{g}_{\mathrm{t}}(\bm{x})/\left\Vert \bm{g}_{\mathrm{t}}(\bm{x})\right\Vert $
is the unit-norm Tx Green vector function and $\widetilde{\bm{P}}_{\mathrm{t},n}\triangleq(\widetilde{\bm{V}}_{n}\widetilde{\bm{V}}_{n}^{\dagger})$
(i.e. the ``noisy'' projector into the right noise subspace). Finally,
a combined version of two modes, named \emph{generalized TR-MUSIC,}
is built as (assuming $M<\min\{N_{T},N_{R}\}$) \cite{Marengo2007}:
\begin{equation}
\mathcal{P}_{\mathrm{tr}}(\bm{x};\widetilde{\bm{U}}_{n},\widetilde{\bm{V}}_{n})\triangleq\mathcal{P}_{\mathrm{t}}(\bm{x};\widetilde{\bm{V}}_{n})+\mathcal{P}_{\mathrm{r}}(\bm{x};\widetilde{\bm{U}}_{n}).\label{eq: TR-MUSIC Tx + Rx Mode}
\end{equation}
Usually, the $M$ largest local maxima of $\mathcal{P}_{\mathrm{r}}(\bm{x};\widetilde{\bm{U}}_{n})^{-1}$,
$\mathcal{P}_{\mathrm{t}}(\bm{x};\widetilde{\bm{V}}_{n})^{-1}$ and
$\mathcal{P}_{\mathrm{tr}}(\bm{x};\widetilde{\bm{U}}_{n},\widetilde{\bm{V}}_{n})^{-1}$
are chosen as the estimates $\{\hat{\bm{x}}_{k}\}_{k=1}^{M}$. Indeed,
it can be shown that Eq.~(\ref{eq: TR-MUSIC Rx Mode}) (resp. Eq.
(\ref{eq: TR-MUSIC Tx Mode})) equals zero when $\bm{x}$ equals one
among $\{\bm{x}_{k}\}_{k=1}^{M}$ in the noise-free case, since when
$\widetilde{\bm{U}}_{n}=\bm{U}_{n}$ (resp. $\widetilde{\bm{V}}_{n}=\bm{V}_{n}$)
this reduces to the eigenvector matrix spanning the left (resp. right)
noise subspace of $\bm{K}(\bm{x}_{1:M},\bm{\tau})$ \cite{Devaney2005}.
Similar conclusions hold for $\mathcal{P}_{\mathrm{tr}}(\bm{x};\widetilde{\bm{U}}_{n},\widetilde{\bm{V}}_{n})$
in a noise-free condition.

\subsection{Review of Results on SVD Perturbation\label{subsec: SVD perturbation preliminaries}}

We consider a rank deficient matrix $\bm{A}\in\mathbb{C}^{R\times T}$
with rank $\delta<\min\{R,T\}$, whose SVD $\bm{A}=\bm{U}\,\bm{\Sigma}\,\bm{V}^{\dagger}$
is rewritten as:
\begin{gather}
\bm{A}=\left(\begin{array}{cc}
\bm{U}_{s} & \bm{U}_{n}\end{array}\right)\left(\begin{array}{cc}
\bm{\Sigma}_{s} & \bm{0}_{\delta\times\check{\delta}}\\
\bm{0}_{\bar{\delta}\times\delta} & \bm{0}_{\bar{\delta}\times\check{\delta}}
\end{array}\right)\left(\begin{array}{c}
\bm{V}_{s}^{\dagger}\\
\bm{V}_{n}^{\dagger}
\end{array}\right)\,,\label{eq: SVD_unpert}
\end{gather}
where $\bar{\delta}\triangleq(R-\delta)$ and $\check{\delta}\triangleq(T-\delta)$,
respectively. Also, $\bm{U}_{s}\in\mathbb{C}^{R\times\delta}$ and
$\bm{V}_{s}\in\mathbb{C}^{T\times\delta}$ (resp. $\bm{U}_{n}\in\mathbb{C}^{R\times\bar{\delta}}$
and $\bm{V}_{n}\in\mathbb{C}^{T\times\check{\delta}}$) denote the
left and right singular vectors of signal (resp. orthogonal) subspaces
in Eq.~(\ref{eq: SVD_unpert}), while $\bm{\Sigma}_{s}\in\mathbb{R}^{\delta\times\delta}$
collects the ($>0$) singular values of the signal subspace. Then,
consider $\widetilde{\bm{A}}=(\bm{A}+\bm{N})$, where $\bm{N}$ is
a perturbing term. Similarly to (\ref{eq: SVD_unpert}), the SVD $\widetilde{\bm{A}}=\widetilde{\bm{U}}\widetilde{\bm{\Sigma}}\widetilde{\bm{V}}^{\dagger}$
is rewritten as
\begin{gather}
\widetilde{\bm{A}}=\left(\begin{array}{cc}
\widetilde{\bm{U}}_{s} & \widetilde{\bm{U}}_{n}\end{array}\right)\left(\begin{array}{cc}
\widetilde{\bm{\Sigma}}_{s} & \bm{0}_{\delta\times\check{\delta}}\\
\bm{0}_{\bar{\delta}\times\delta} & \widetilde{\bm{\Sigma}}_{n}
\end{array}\right)\left(\begin{array}{c}
\widetilde{\bm{V}}_{s}^{\dagger}\\
\widetilde{\bm{V}}_{n}^{\dagger}
\end{array}\right)\,,\label{eq: SVD_perturb}
\end{gather}
showing the effect of $\bm{N}$ on the spectral representation\footnote{Indeed, as opposed to Eq. (\ref{eq: SVD_unpert}), $\widetilde{\bm{A}}$
may be full-rank in general.} of $\widetilde{\bm{A}}$, highlighting the change of the left and
right principal directions. We are here concerned with the perturbations
pertaining to $\widetilde{\bm{U}}_{n}$ and $\widetilde{\bm{V}}_{n}$,
stressed as $\widetilde{\bm{U}}_{n}=\bm{U}_{n}+\bm{\Delta U}_{n}$
and $\widetilde{\bm{V}}_{n}=\bm{V}_{n}+\bm{\Delta V}_{n}$, where
$\bm{\Delta}(\cdot)$ terms are generally complicated functions of
$\bm{N}$. However, when $\bm{N}$ has a ``small magnitude'' compared
to $\bm{A}$ (see \cite{Stewart1973}), a 1st-order perturbation (i.e.
$\bm{\Delta}(\cdot)$ are approximated as linear with $\bm{N}$),
will be accurate \cite{Li1993}. The key result is that perturbed
orthogonal left subspace $\widetilde{\bm{U}}_{n}$ (resp. right subspace
$\widetilde{\bm{V}}_{n}$) is spanned by $\bm{U}_{n}+\bm{U}_{s}\bm{B}$
(resp. $\bm{V}_{n}+\bm{V}_{s}\bar{\bm{B}}$), where norm (any sub-multiplicative
one, such as $\ell_{2}$ or $\left\Vert \cdot\right\Vert _{F}$ norm)
of $\bm{B}$ (resp. $\bar{\bm{B}}$) is of the same order of that
of $\bm{N}$. Intuitively, a \emph{small perturbation} is observed
at \emph{high-SNR}. The expressions for $\bm{\Delta U}_{n}$ and $\bm{\Delta V}_{n}$,
at 1st-order, are\footnote{We notice that in obtaining Eq. (\ref{eq: Delta_Un-Vn}), ``in-space''
perturbations (e.g. the contribution to $\bm{\Delta U}_{n}$ depending
on $\bm{U}_{n}$) are not considered, though they have been shown
to be linear with $\bm{N}$ (and thus \emph{not negligible} at first-order)
\cite{Liu2008}. The reason is that these terms do not affect performance
analysis of TR-MUSIC null-spectrum when evaluated at scatterers positions
$\{\bm{x}_{k}\}_{k=1}^{M}$, due to the null spectrum orthogonality
property.} \cite{Liu2008}:
\begin{gather}
\bm{\Delta U}_{n}=-(\bm{A}^{-})^{\dagger}\,\bm{N}^{\dagger}\,\bm{U}_{n};\quad\bm{\Delta V}_{n}=-(\bm{A}^{-})\,\bm{N}\,\bm{V}_{n};\label{eq: Delta_Un-Vn}
\end{gather}
where we have exploited $\bm{A}^{-}=\bm{V}_{s}\,\bm{\Sigma}_{s}^{-1}\,\bm{U}_{s}^{\dagger}$
\cite{Bernstein2009}.

\section{Null-spectrum analysis \label{sec: Performance analysis}}

First, we observe that the null spectrums at scatterer positions $\mathcal{P}_{\mathrm{r}}(\bm{x}_{k};\widetilde{\bm{U}}_{n})$,
$\mathcal{P}_{\mathrm{t}}(\bm{x}_{k};\widetilde{\bm{V}}_{n})$ and
$\mathcal{P}_{\mathrm{tr}}(\bm{x}_{k};\widetilde{\bm{U}}_{n},\widetilde{\bm{V}}_{n})$,
$k\in\{1,\ldots,M\}$, in Eqs. (\ref{eq: TR-MUSIC Rx Mode}), (\ref{eq: TR-MUSIC Tx Mode})
and (\ref{eq: TR-MUSIC Tx + Rx Mode}) can be simplified, using $\widetilde{\bm{U}}_{n}=\bm{U}_{n}+\bm{\Delta U}_{n}$
and $\widetilde{\bm{V}}_{n}=\bm{V}_{n}+\bm{\Delta V}_{n}$ and exploiting
the properties\footnote{Such conditions directly follow from orthogonality between left (resp.
right) signal and orthogonal subspaces $\bm{U}_{s}$ and $\bm{U}_{n}$
(resp. $\bm{V}_{s}$ and $\bm{V}_{n}$).} $\bm{U}_{n}^{\dagger}\,\bar{\bm{g}}_{\mathrm{r}}(\bm{x}_{k})=\bm{0}_{N_{\mathrm{Rdof}}}$
and $\bm{V}_{n}^{\dagger}\bar{\bm{g}}_{\mathrm{t}}^{*}(\bm{x}_{k})=\bm{0}_{N_{\mathrm{Tdof}}}$,
as
\begin{equation}
\mathcal{P}_{\mathrm{r}}(\bm{x}_{k};\widetilde{\bm{U}}_{n})=\left\Vert \bm{\xi}_{\mathrm{r,}k}\right\Vert ^{2}\,,\quad\mathcal{P}_{\mathrm{t}}(\bm{x}_{k};\widetilde{\bm{V}}_{n})=\left\Vert \bm{\xi}_{\mathrm{t,}k}\right\Vert ^{2},\label{eq: TR-MUSIC quadratic form Rx and Tx modes}
\end{equation}
where $\bm{\xi}_{\mathrm{r,}k}\triangleq\bm{\Delta U}_{n}^{\dagger}\,\bar{\bm{g}}_{\mathrm{r}}(\bm{x}_{k})\in\mathbb{C}^{N_{\mathrm{Rdof}}\times1}$
and $\bm{\xi}_{\mathrm{t,}k}\triangleq\bm{\Delta V}_{n}^{\dagger}\,\bar{\bm{g}}_{t}^{*}(\bm{x}_{k})\in\mathbb{C}^{N_{\mathrm{Tdof}}\times1}$,
respectively. Similarly, 
\begin{equation}
\mathcal{P}_{\mathrm{tr}}(\bm{x}_{k};\widetilde{\bm{U}}_{n},\widetilde{\bm{V}}_{n})=\left\Vert \bm{\xi}_{\mathrm{t,}k}\right\Vert ^{2}+\left\Vert \bm{\xi}_{\mathrm{r,}k}\right\Vert ^{2}=\left\Vert \bm{\xi}_{k}\right\Vert ^{2},\label{eq: TR-MUSIC quadratic form Rx+Tx modes}
\end{equation}
where $\bm{\xi}_{k}\triangleq\begin{bmatrix}\bm{\xi}_{\mathrm{r,}k}^{T} & \bm{\xi}_{\mathrm{t,}k}^{T}\end{bmatrix}^{T}\in\mathbb{C}^{N_{\mathrm{dof}}\times1}$.
Thus, to characterize $\mathcal{P}_{\mathrm{r}}(\bm{x}_{k};\widetilde{\bm{U}}_{n})$,
$\mathcal{P}_{\mathrm{t}}(\bm{x}_{k};\widetilde{\bm{V}}_{n})$ and
$\mathcal{P}_{\mathrm{tr}}(\bm{x}_{k};\widetilde{\bm{U}}_{n},\widetilde{\bm{V}}_{n})$,
it suffices to study the random vector $\bm{\xi}_{k}$. Indeed, the
marginal pdfs of $\bm{\xi}_{\mathrm{r,}k}$ and $\bm{\xi}_{\mathrm{t,}k}$
are easily drawn from that of $\bm{\xi}_{k}$. As a byproduct, $\bm{\xi}_{k}$
definition also allows an elegant and simpler MSE analysis with respect
to \cite{Ciuonzo2015}, as it can be shown that the position-error
of the estimates with Tx mode ($\bm{\Delta}\bm{x}_{T,k}$), Rx mode
($\bm{\Delta}\bm{x}_{R,k}$) and generalized ($\bm{\Delta}\bm{x}_{TR,k}$)
TR-MUSIC can be expressed as $\bm{\Delta}\bm{x}_{T,k}\approx-\bm{\Gamma}_{T,k}^{-1}\Re\{\bm{J}_{T,k}^{T}\,\bm{V}_{n}\,\bm{\xi}_{\mathrm{t,}k}\}$,
$\bm{\Delta}\bm{x}_{R,k}\approx-\bm{\Gamma}_{R,k}^{-1}\Re\{\bm{J}_{R,k}^{\dagger}\,\bm{U}_{n}\,\bm{\xi}_{\mathrm{r,}k}\}$
and $\bm{\Delta}\bm{x}_{TR,k}\approx-\bm{\Gamma}_{TR,k}^{-1}\Re\{[\begin{array}{cc}
(\bm{J}_{R,k}^{\dagger}\,\bm{U}_{n}) & (\bm{J}_{T,k}^{T}\,\bm{V}_{n})\end{array}]\bm{\xi}_{k}\}$, respectively, where $\bm{J}_{T,k}$, $\bm{J}_{R,k}$, $\bm{\Gamma}_{T,k}$,
$\bm{\Gamma}_{R,k}$ and $\bm{\Gamma}_{TR,k}$ are suitably defined
\emph{known} matrices (see \cite{Ciuonzo2015}). Clearly, finding
the exact pdf of $\bm{\xi}_{k}$ is hard, as $\bm{\Delta U}_{n}$
and $\bm{\Delta V}_{n}$ are generally complicated functions of the
unknown \emph{perturbing }matrix $\bm{W}$.

However, $\bm{\Delta U}_{n}$ and $\bm{\Delta V}_{n}$ assume a (tractable)
closed form with a 1st-order approximation (see Eq. (\ref{eq: Delta_Un-Vn})).
This approximation holds tightly at high-SNR, as $\bm{W}$ will be
statistically ``small'' compared to noise-free MDM $\bm{K}(\bm{x}_{1:M},\bm{\tau})$.
Hence, at high-SNR, $\bm{\xi}_{k}$ is (approximately) expressed in
terms of $\bm{W}$ as:
\begin{equation}
\bm{\xi}_{k}=\begin{bmatrix}\bm{\xi}_{\mathrm{r,}k}\\
\bm{\xi}_{\mathrm{t,}k}
\end{bmatrix}\approx\begin{bmatrix}-\bm{U}_{n}^{\dagger}\,\bm{W}\,\bm{t}_{\mathrm{r},k}\\
-\bm{V}_{n}^{\dagger}\,\bm{W}^{\dagger}\,\bm{t}_{\mathrm{t},k}
\end{bmatrix}\,,\label{eq: Xi_k approximated}
\end{equation}
where $\bm{t}_{\mathrm{r},k}\triangleq\bm{K}^{-}(\bm{x}_{1:M},\bm{\tau})\,\bar{\bm{g}}_{\mathrm{r}}(\bm{x}_{k})\in\mathbb{C}^{N_{T}\times1}$
and $\bm{t}_{\mathrm{t},k}\triangleq\bm{K}^{-}(\bm{x}_{1:M},\bm{\tau})^{\dagger}\,\bar{\bm{g}}_{\mathrm{t}}^{\text{*}}(\bm{x}_{k})\in\mathbb{C}^{N_{R}\times1}$
are \emph{deterministic.} Since the vector $\bm{\xi}_{k}$ is linear\footnote{In the following of the letter we will implicitly mean that the results
hold ``approximately'' in the high-SNR regime.} with the noise matrix $\bm{W}$, it will be Gaussian distributed;
thus we only need to evaluate its moments up to the 2nd order to characterize
it completely. Hereinafter we only sketch the main steps and provide
the detailed proof as supplementary material. First, the mean vector
$\mathbb{E}\{\begin{bmatrix}\bm{\xi}_{\mathrm{r,}k}^{T} & \bm{\xi}_{\mathrm{t,}k}^{T}\end{bmatrix}^{T}\}=\bm{0}_{N_{\mathrm{dof}}}$,
exploiting $\mathbb{E}\left\{ \bm{W}\right\} =\bm{0}_{N_{R}\times N_{T}}$.
Secondly, the \emph{covariance matrix} $\bm{\Xi}_{k}\triangleq\mathbb{E}\{\bm{\xi}_{k}\bm{\xi}_{k}^{\dagger}\}$
(since $\mathbb{E}\{\bm{\xi}_{k}\}=\bm{0}_{N_{\mathrm{dof}}}$) is
given in closed-form as:
\begin{gather}
\bm{\Xi}_{k}=\begin{bmatrix}\sigma_{w}^{2}\,\left\Vert \bm{t}_{\mathrm{r},k}\right\Vert ^{2}\,\bm{I}_{N_{\mathrm{Rdof}}} & \bm{0}_{N_{\mathrm{Rdof}}\times N_{\mathrm{Tdof}}}\\
\bm{0}_{N_{\mathrm{Tdof}}\times N_{\mathrm{Rdof}}} & \sigma_{w}^{2}\,\left\Vert \bm{t}_{\mathrm{t},k}\right\Vert ^{2}\,\bm{I}_{N_{\mathrm{Tdof}}}
\end{bmatrix}\,.\label{eq: cov xi_k final}
\end{gather}
The above result is based on circularity of the entries of $\bm{W}$,
along with their mutual independence. Thirdly, aiming at completing
the statistical characterization, we evaluate the \emph{pseudo-covariance
matrix} $\bm{\Psi}_{k}\triangleq\mathbb{E}\{\bm{\xi}_{k}\bm{\xi}_{k}^{T}\}$
(since $\mathbb{E}\{\bm{\xi}_{k}\}=\bm{0}_{N_{\mathrm{dof}}}$), whose
closed-form is $\bm{\Psi}_{k}=\bm{0}_{N_{\mathrm{dof}}\times N_{\mathrm{dof}}}$.
The latter result is based on circularity of the entries of $\bm{W}$,
along with their mutual independence and exploiting the results $\bm{V}_{n}^{\dagger}\,\bm{t}_{\mathrm{r},k}=\bm{0}_{N_{\mathrm{Tdof}}}$
and $\bm{U}_{n}^{\dagger}\,\bm{t}_{\mathrm{t,k}}=\bm{0}_{N_{\mathrm{Rdof}}}$,
arising from subspaces orthogonality $\bm{V}_{n}^{\dagger}\bm{V}_{s}=\bm{0}_{N_{\mathrm{Tdof}}\times M}$
and $\bm{U}_{n}^{\dagger}\bm{U}_{s}=\bm{0}_{N_{\mathrm{Rdof}}\times M}$.

Therefore, in summary $\bm{\xi}_{k}\sim\mathcal{N}_{\mathbb{C}}\left(\bm{0}_{N_{\mathrm{dof}}},\,\bm{\Xi}_{k}\right)$,
i.e. a \emph{proper} complex Gaussian vector \cite{Schreier2010}.
Similarly, it is readily inferred that $\bm{\xi}_{\mathrm{r},k}\sim\,\mathcal{N}_{\mathbb{C}}(\bm{0}_{N_{\mathrm{Rdof}}},\,\sigma_{w}^{2}\,\left\Vert \bm{t}_{\mathrm{r},k}\right\Vert ^{2}\,\bm{I}_{N_{\mathrm{Rdof}}})$
and $\bm{\xi}_{\mathrm{t},k}\sim\,\mathcal{N}_{\mathbb{C}}(\bm{0}_{N_{\mathrm{Tdof}}},\,\sigma_{w}^{2}\,\left\Vert \bm{t}_{\mathrm{t},k}\right\Vert ^{2}\,\bm{I}_{N_{\mathrm{Tdof}}})$,
respectively, i.e. they are\emph{ independent} \emph{proper} Gaussian
vectors. Clearly, since $\bm{\xi}_{\mathrm{r,}k}$ and $\bm{\xi}_{t,k}$
have zero mean and scaled-identity covariance, the corresponding \emph{variance-normalized}
energies $\left\Vert \bm{\xi}_{\mathrm{r},k}\right\Vert ^{2}/(\sigma_{w}^{2}\left\Vert \bm{t}_{\mathrm{r},k}\right\Vert ^{2})\sim\mathcal{C\chi}_{N_{\mathrm{Rdof}}}^{2}$
and $\left\Vert \bm{\xi}_{\mathrm{t},k}\right\Vert ^{2}/(\sigma_{w}^{2}\left\Vert \bm{t}_{\mathrm{t},k}\right\Vert ^{2})\sim\mathcal{C\chi}_{N_{\mathrm{Tdof}}}^{2}$,
respectively (i.e. they are chi-square distributed). Interestingly
these DOFs coincide with those available for TR-MUSIC localization
through Rx and Tx modes, respectively.

Based on these considerations, the means of the null-spectrum for
Tx and Rx modes are $\mathbb{E}\{\left\Vert \bm{\xi}_{\mathrm{r},k}\right\Vert ^{2}\}=\sigma_{w}^{2}\,\left\Vert \bm{t}_{\mathrm{r},k}\right\Vert ^{2}\,N_{\mathrm{Rdof}}$
and $\mathbb{E}\{\left\Vert \bm{\xi}_{\mathrm{t},k}\right\Vert ^{2}\}=\sigma_{w}^{2}\,\left\Vert \bm{t}_{\mathrm{t},k}\right\Vert ^{2}\,N_{\mathrm{Tdof}}$,
respectively, whereas for generalized null-spectrum $\mathbb{E}\{\left\Vert \bm{\xi}_{k}\right\Vert ^{2}\}=\mathbb{E}\{\left\Vert \bm{\xi}_{\mathrm{r},k}\right\Vert ^{2}\}+\mathbb{E}\{\left\Vert \bm{\xi}_{\mathrm{t},k}\right\Vert ^{2}\}$
(by linearity). By similar reasoning, the variances for Tx and Rx
modes are given by $\mathrm{var}\{\left\Vert \bm{\xi}_{\mathrm{r},k}\right\Vert ^{2}\}=\sigma_{w}^{4}\,\left\Vert \bm{t}_{\mathrm{r},k}\right\Vert ^{4}\,N_{\mathrm{Rdof}}$
and $\mathrm{var}\{\left\Vert \bm{\xi}_{\mathrm{t},k}\right\Vert ^{2}\}=\sigma_{w}^{4}\,\left\Vert \bm{t}_{\mathrm{t},k}\right\Vert ^{4}\,N_{\mathrm{Tdof}}$,
respectively, whereas for the generalized null-spectrum $\mathrm{var}\{\left\Vert \bm{\xi}_{k}\right\Vert ^{2}\}=\mathrm{var}\{\left\Vert \bm{\xi}_{\mathrm{r},k}\right\Vert ^{2}\}+\mathrm{var}\{\left\Vert \bm{\xi}_{\mathrm{t},k}\right\Vert ^{2}\}$
(by independence of $\bm{\xi}_{\mathrm{r,}k}$ and $\bm{\xi}_{\mathrm{t,}k}$).

Hence, once we have obtained the mean and the variance of $\mathcal{P}_{\mathrm{r}}(\bm{x}_{k};\widetilde{\bm{U}}_{n})$,
$\mathcal{P}_{\mathrm{t}}(\bm{x}_{k};\widetilde{\bm{V}}_{n})$ and
$\mathcal{P}_{\mathrm{tr}}(\bm{x}_{k};\widetilde{\bm{U}}_{n},\widetilde{\bm{V}}_{n})$,
respectively, we can consider the \emph{Normalized Standard Deviation}
(NSD), generically defined as
\begin{equation}
\mathrm{NSD}_{k}\triangleq\,\sqrt{\mathrm{var}\{\mathcal{P}(\bm{x}_{k};\,\cdot)}\}\,/\,\mathbb{E}\{\mathcal{P}(\bm{x}_{k};\,\cdot)\}.
\end{equation}
Clearly, the lower the NSD, the higher the null-spectrum stability
at $\bm{x}_{k}$ \cite{Choi1993}. For \emph{Rx} and \emph{Tx} \emph{modes}
it follows that $\mathrm{NSD}_{\mathrm{r},k}=1/\sqrt{N_{\mathrm{Rdof}}}$
and $\mathrm{NSD}_{\mathrm{t},k}=1/\sqrt{N_{\mathrm{Tdof}}}$, respectively.
It is apparent that in both cases the NSD \emph{does not depend} (at
high SNR) on the scatterers and measurement setup, as well as $\sigma_{w}^{2}$,
but only on the (complex) DOFs, being equal to $N_{\mathrm{Rdof}}$
and $N_{\mathrm{Tdof}}$, respectively. Thus, the NSD becomes (asymptotically)
small only when the number of scatterers is few compared to the Tx
(resp. Rx) elements of the array. Those results are analogous to the
case of MUSIC null-spectrum for DOA, whose NSD depends on the DOFs,
namely the difference between the (Rx) array size and the number of
sources \cite{Choi1993}. Differently, the NSD for generalized null
spectrum equals 
\begin{equation}
\mathrm{NSD}_{k}=\frac{\sqrt{\left\Vert \bm{t}_{\mathrm{r},k}\right\Vert ^{4}N_{\mathrm{Rdof}}+\left\Vert \bm{t}_{\mathrm{t},k}\right\Vert ^{4}N_{\mathrm{Tdof}}}}{\left\Vert \bm{t}_{\mathrm{r},k}\right\Vert ^{2}N_{\mathrm{Rdof}}+\left\Vert \bm{t}_{\mathrm{t},k}\right\Vert ^{2}N_{\mathrm{Tdof}}}\,.\label{eq: NSD tx_rx spectrum}
\end{equation}
Eq. (\ref{eq: NSD tx_rx spectrum}) underlines ($i$) a clear dependence
of generalized null-spectrum NSD on scatterers and measurement setup
and ($ii$) independence from the noise level $\sigma_{w}^{2}$. Also,
it is apparent that when $\left\Vert \bm{t}_{\mathrm{r},k}\right\Vert \approx0$
(resp. $\left\Vert \bm{t}_{\mathrm{t},k}\right\Vert \approx0$) the
expression reduces to $\mathrm{NSD}_{k}\approx1/\sqrt{N_{\mathrm{Tdof}}}$
(resp. $\mathrm{NSD}_{k}\approx1/\sqrt{N_{\mathrm{Rdof}}}$), i.e.
the NSD is \emph{dominated} by Tx (resp. Rx) mode stability. Finally,
the same equation is exploited to obtain the conditions ensuring that
generalized spectrum is ``more stable'' than Tx and Rx modes ($\mathrm{NSD}_{k}\leq\mathrm{NSD}_{\mathrm{t},k}$
and $\mathrm{NSD}_{k}\leq\mathrm{NSD}_{\mathrm{r},k}$, respectively),
expressed as the pair of inequalities
\begin{gather}
\begin{cases}
\frac{1}{2}\left[1-N_{\mathrm{Rdof}}/N_{\mathrm{Tdof}}\right]\leq\left(\left\Vert \bm{t}_{\mathrm{t},k}\right\Vert /\left\Vert \bm{t}_{\mathrm{r},k}\right\Vert \right)^{2} & (\mathrm{Tx})\\
\frac{1}{2}\left[1-N_{\mathrm{Tdof}}/N_{\mathrm{Rdof}}\right]\leq\left(\left\Vert \bm{t}_{\mathrm{r},k}\right\Vert /\left\Vert \bm{t}_{\mathrm{t},k}\right\Vert \right)^{2} & (\mathrm{Rx})
\end{cases}\label{eq: pair inequalities}
\end{gather}
Clearly, when $N_{R}>N_{T}$ (resp. $N_{T}>N_{R}$) the inequality
regarding the Tx (resp. Rx) mode is always verified as the left-hand
side is always negative. Also, in the special case $N_{T}=N_{R}$
the left-hand side is always zero for both inequalities. 
\begin{figure}[t]
\centering{}\includegraphics[width=0.9\columnwidth]{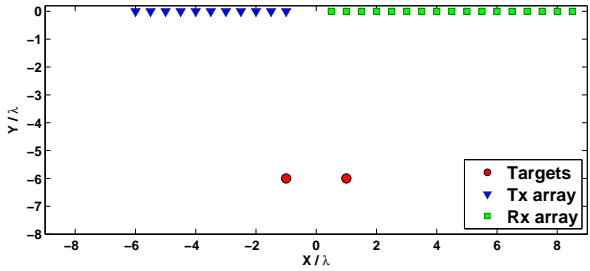}\caption{Geometry for the considered imaging problem in 2-D space.\label{fig: Setup non-colocated}}
\end{figure}
\begin{figure}[t]
\centering{}\includegraphics[width=1\columnwidth]{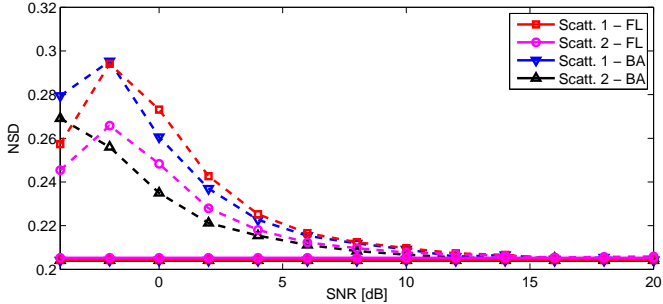}\caption{ NSD (generalized null spectrum) vs. SNR; theoretical (Eq. (\ref{eq: NSD tx_rx spectrum}),
solid lines) vs. MC-based (dashed lines) performance.\label{fig: 1st exp mean vs SNR}}
\end{figure}
\begin{figure}[t]
\centering{}\includegraphics[width=1\columnwidth]{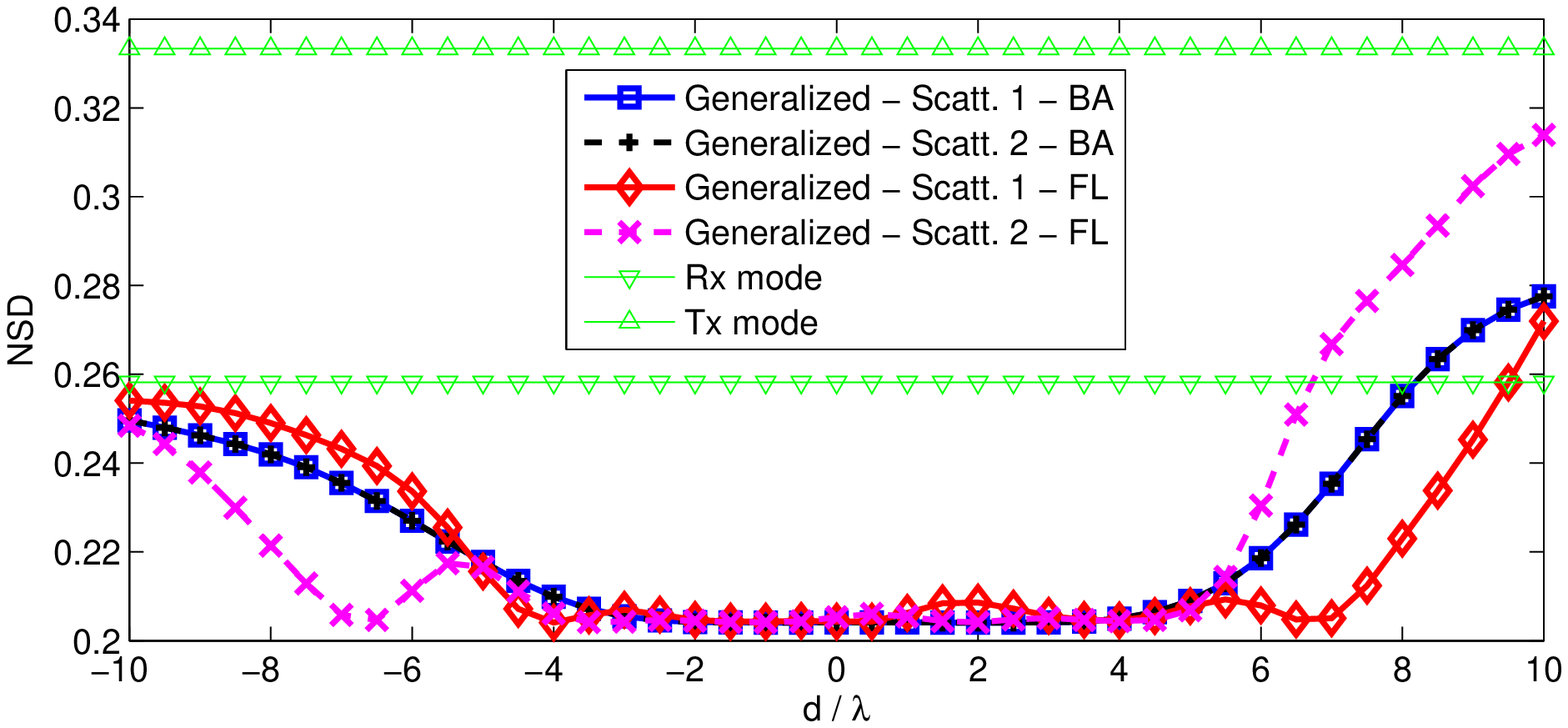}\caption{Theoretical NSD vs. scatterers rigid shift $d$; two targets located
at $(\bm{x}_{1}/\lambda)=\left[\protect\begin{array}{cc}
(-1-d) & -6\protect\end{array}\right]^{T}$ and $(\bm{x}_{2}/\lambda)=\left[\protect\begin{array}{cc}
(1-d) & -6\protect\end{array}\right]^{T}$.\label{fig: NSD vs displacement}}
\end{figure}

\section{Numerical results\label{sec: Numerical Results}}

In this section we confirm our findings through simulations, focusing
on 2-D localization, with Green function\footnote{We discard the irrelevant constant term $j/4$.}
being $\mathcal{G}(\bm{x}',\bm{x})=H_{0}^{(1)}\left(\kappa\left\Vert \bm{x}'-\bm{x}\right\Vert \right)$.
Here $H_{n}^{(1)}(\cdot)$ and $\kappa=2\pi/\lambda$ denote the $n$th
order \emph{Hankel} function of the 1st kind and the wavenumber ($\lambda$
is the wavelength), respectively. First, we consider a setup with
$\lambda/2$-spaced Tx/Rx arrays ($N_{T}=11$ and $N_{R}=17$, respectively,
see Fig.~\ref{fig: Setup non-colocated}). Secondly, to quantify
the level of multiple scattering (as in \cite{Marengo2007}) we define
the index $\eta\triangleq\left\Vert \bm{K}_{\mathrm{f}}(\bm{x}_{1:M},\bm{\tau})-\bm{K}_{\mathrm{b}}(\bm{x}_{1:M},\bm{\tau})\right\Vert _{F}/\left\Vert \bm{K}_{\mathrm{b}}(\bm{x}_{1:M},\bm{\tau})\right\Vert _{F},$
where $\bm{K}_{\mathrm{b}}(\bm{x}_{1:M},\bm{\tau})$ and $\bm{K}_{\mathrm{f}}(\bm{x}_{1:M},\bm{\tau})$
denote the MDMs generated via BA and FL models, respectively. Finally,
for simplicity we consider $M=2$ targets located at $(\bm{x}_{1}/\lambda)=\left[\begin{array}{cc}
-1 & -6\end{array}\right]^{T}$ and $(\bm{x}_{2}/\lambda)=\left[\begin{array}{cc}
+1 & -6\end{array}\right]^{T}$ and having scattering coefficients $\bm{\tau}=\left[\begin{array}{cc}
3 & 4\end{array}\right]^{T}$; thus $\eta=(0.7445)$.

Then, we compare the asymptotic NSD (Eq.~(\ref{eq: NSD tx_rx spectrum}),
solid lines) with the true ones obtained via Monte Carlo (MC) simulation
(dashed lines, $10^{5}$ runs), focusing only on the generalized null-spectrum
for brevity. To this end, Fig.~\ref{fig: 1st exp mean vs SNR} depicts
the null-spectrum NSD vs. SNR for the two targets being considered,
both for FL and BA models. It is apparent that, as the SNR increases,
the theoretical results tightly approximate the MC-based ones, with
approximations deemed accurate above $\mathrm{SNR}\approx10\,\mathrm{dB}$.
Differently, in Fig.~\ref{fig: NSD vs displacement}, we plot the
asymptotic NSD of the three TR-MUSIC variants vs. $d$, where $(\bm{x}_{1}/\lambda)=\left[\begin{array}{cc}
(-1-d) & -6\end{array}\right]^{T}$ and $(\bm{x}_{2}/\lambda)=\left[\begin{array}{cc}
(1-d) & -6\end{array}\right]^{T}$ (i.e. a rigid shift of the two scatterers), in order to investigate
the potentially improved asymptotic stability (viz. NSD) of the generalized
spectrum in comparison to Tx and Rx modes. It is apparent that the
gain is significant when $d\in(-5,5)$, while outside this interval
the NSD expression is either dominated by Tx or Rx mode, which for
the present case $\mathrm{NSD}_{\mathrm{t},k}=1/\sqrt{11-2}\approx0.33$
and $\mathrm{NSD}_{\mathrm{r},k}=1/\sqrt{17-2}\approx0.26$, with
the generalized NSD never above that of $\mathrm{NSD}_{\mathrm{t},k}$
(as dictated from Eq. (\ref{eq: pair inequalities})).

\section{Conclusions\label{sec: Conclusions}}

We provided an asymptotic (high-SNR) analysis of TR-MUSIC null-spectrum
in a non-colocated multistatic setup, by taking advantage of the 1st-order
perturbation of the SVD of the MDM. Three different variants of TR-MUSIC
were analyzed (i.e. Tx mode, Rx mode and generalized), based on the
characterization of a certain complex-valued Gaussian vector. This
allowed to obtain the asymptotic NSD (a measure of null-spectrum stability)
for all the three imaging procedures. While similar results as the
DOA setup were obtained for Tx and Rx modes, it was shown a clear
dependence of generalized null-spectrum NSD on the scatterer and measurement
setup. Finally, its potential stability advantage was investigated
in comparison to Tx and Rx modes. Future works will analyze mutual
coupling, antenna pattern and polarization effects \cite{Agarwal2008,Solimene2014},
and propagation in inhomogeneous (random) media \cite{Fouda2014}.
\pagebreak{}

\bibliographystyle{IEEEtran}
\bibliography{IEEEabrv,imaging_refs}

\begin{thebibliography}{10}
\providecommand{\url}[1]{#1}
\csname url@samestyle\endcsname
\providecommand{\newblock}{\relax}
\providecommand{\bibinfo}[2]{#2}
\providecommand{\BIBentrySTDinterwordspacing}{\spaceskip=0pt\relax}
\providecommand{\BIBentryALTinterwordstretchfactor}{4}
\providecommand{\BIBentryALTinterwordspacing}{\spaceskip=\fontdimen2\font plus
\BIBentryALTinterwordstretchfactor\fontdimen3\font minus
  \fontdimen4\font\relax}
\providecommand{\BIBforeignlanguage}[2]{{%
\expandafter\ifx\csname l@#1\endcsname\relax
\typeout{** WARNING: IEEEtran.bst: No hyphenation pattern has been}%
\typeout{** loaded for the language `#1'. Using the pattern for}%
\typeout{** the default language instead.}%
\else
\language=\csname l@#1\endcsname
\fi
#2}}
\providecommand{\BIBdecl}{\relax}
\BIBdecl

\bibitem{Fink1993}
M.~Fink, ``Time-reversal mirrors,'' \emph{Journal of Physics D: Applied
  Physics}, vol.~26, no.~9, p. 1333, 1993.

\bibitem{Cassereau1992}
D.~Cassereau and M.~Fink, ``Time-reversal of ultrasonic fields. {III.} theory
  of the closed time-reversal cavity,'' \emph{{IEEE} Trans. Ultrason.,
  Ferroelectr., Freq. Control}, vol.~39, no.~5, pp. 579--592, Sep. 1992.

\bibitem{Micolau2003}
G.~Micolau, M.~Saillard, and P.~Borderies, ``{DORT} method as applied to
  ultrawideband signals for detection of buried objects,'' \emph{IEEE Trans.
  Geosci. Remote Sens.}, vol.~41, no.~8, pp. 1813--1820, Aug. 2003.

\bibitem{Li2010}
L.~Li, W.~Zhang, and F.~Li, ``A novel autofocusing approach for real-time
  through-wall imaging under unknown wall characteristics,'' \emph{{IEEE}
  Trans. Geosci. Remote Sens.}, vol.~48, no.~1, pp. 423--431, Jan. 2010.

\bibitem{Hossain2013}
M.~D. Hossain, A.~S. Mohan, and M.~J. Abedin, ``Beamspace time-reversal
  microwave imaging for breast cancer detection,'' \emph{{IEEE} Antennas
  Wireless Propag. Lett.}, vol.~12, pp. 241--244, 2013.

\bibitem{Prada1996}
C.~Prada, S.~Manneville, D.~Spoliansky, and M.~Fink, ``Decomposition of the
  time reversal operator: Detection and selective focusing on two scatterers,''
  \emph{The Journal of the Acoustical Society of America}, vol.~99, no.~4, pp.
  2067--2076, 1996.

\bibitem{Devaney2005}
A.~J. Devaney, ``Time reversal imaging of obscured targets from multistatic
  data,'' \emph{{IEEE} Trans. Antennas Propag.}, vol.~53, no.~5, pp.
  1600--1610, May 2005.

\bibitem{Marengo2007}
E.~A. Marengo and F.~K. Gruber, ``Subspace-based localization and inverse
  scattering of multiply scattering point targets,'' \emph{EURASIP Journal on
  Advances in Signal Processing}, pp. 1--16, 2007.

\bibitem{Marengo2007a}
E.~A. Marengo, F.~K. Gruber, and F.~Simonetti, ``Time-reversal {MUSIC} imaging
  of extended targets,'' \emph{{IEEE} Trans. Image Process.}, vol.~16, no.~8,
  pp. 1967--1984, Aug. 2007.

\bibitem{Schmidt1986}
R.~Schmidt, ``Multiple emitter location and signal parameter estimation,''
  \emph{{IEEE} Trans. Antennas Propag.}, vol.~34, no.~3, pp. 276--280, 1986.

\bibitem{Kaveh1986}
M.~Kaveh and A.~Barabell, ``The statistical performance of the {MUSIC} and the
  minimum-norm algorithms in resolving plane waves in noise,'' \emph{{IEEE}
  Trans. Acoust., Speech, Signal Process.}, vol.~34, no.~2, pp. 331--341, 1986.

\bibitem{Friedlander1990}
B.~Friedlander, ``A sensitivity analysis of the {MUSIC} algorithm,''
  \emph{{IEEE} Trans. Acoust., Speech, Signal Process.}, vol.~38, no.~10, pp.
  1740--1751, Oct. 1990.

\bibitem{Porat1988}
B.~Porat and B.~Friedlander, ``Analysis of the asymptotic relative efficiency
  of the {MUSIC} algorithm,'' \emph{{IEEE} Trans. Acoust., Speech, Signal
  Process.}, vol.~36, no.~4, pp. 532--544, Apr. 1988.

\bibitem{Stoica1989}
P.~Stoica and A.~Nehorai, ``{MUSIC}, maximum likelihood, and {Cramér-Rao}
  bound,'' \emph{{IEEE} Trans. Acoust., Speech, Signal Process.}, vol.~37,
  no.~5, pp. 720--741, May 1989.

\bibitem{Li1990}
F.~Li and R.~J. Vaccaro, ``Analysis of min-norm and {MUSIC} with arbitrary
  array geometry,'' \emph{{IEEE} Trans. Aerosp. Electron. Syst.}, vol.~26,
  no.~6, pp. 976--985, 1990.

\bibitem{Swindlehurst1992}
A.~L. Swindlehurst and T.~Kailath, ``A performance analysis of subspace-based
  methods in the presence of model errors, {Part I:} the {MUSIC} algorithm,''
  \emph{{IEEE} Trans. Signal Process.}, vol.~40, no.~7, pp. 1758--1774, Jul.
  1992.

\bibitem{Ferreol2006}
A.~Ferr{\'e}ol, P.~Larzabal, and M.~Viberg, ``On the asymptotic performance
  analysis of subspace {DOA} estimation in the presence of modeling errors:
  case of {MUSIC},'' \emph{{IEEE} Trans. Signal Process.}, vol.~54, no.~3, pp.
  907--920, Mar. 2006.

\bibitem{Ferreol2008}
------, ``On the resolution probability of {MUSIC} in presence of modeling
  errors,'' \emph{{IEEE} Trans. Signal Process.}, vol.~56, no.~5, pp.
  1945--1953, May 2008.

\bibitem{Ferreol2010}
------, ``Statistical analysis of the {MUSIC} algorithm in the presence of
  modeling errors, taking into account the resolution probability,''
  \emph{{IEEE} Trans. Signal Process.}, vol.~58, no.~8, pp. 4156--4166, Aug.
  2010.

\bibitem{Ciuonzo2014}
D.~Ciuonzo, G.~Romano, and R.~Solimene, ``On {MSE} performance of time-reversal
  {MUSIC},'' in \emph{IEEE 8th Sensor Array and Multichannel Signal Processing
  Workshop (SAM)}, Jun. 2014, pp. 13--16.

\bibitem{Ciuonzo2015}
------, ``Performance analysis of time-reversal {MUSIC},'' \emph{{IEEE} Trans.
  Signal Process.}, vol.~63, no.~10, pp. 2650--2662, 2015.

\bibitem{Shi2007a}
G.~Shi and A.~Nehorai, ``{Cramér-Rao} bound analysis on multiple scattering in
  multistatic point-scatterer estimation,'' \emph{{IEEE} Trans. Signal
  Process.}, vol.~55, no.~6, pp. 2840--2850, Jun. 2007.

\bibitem{Choi1993}
J.~Choi and I.~Song, ``Asymptotic distribution of the {MUSIC} null spectrum,''
  \emph{{IEEE} Trans. Signal Process.}, vol.~41, no.~2, pp. 985--988, 1993.

\bibitem{Li1993}
F.~Li, H.~Liu, and R.~J. Vaccaro, ``Performance analysis for {DOA} estimation
  algorithms: unification, simplification, and observations,'' \emph{{IEEE}
  Trans. Aerosp. Electron. Syst.}, vol.~29, no.~4, pp. 1170--1184, 1993.

\bibitem{CiuonzoSubmitted2016}
D.~Ciuonzo and {P. Salvo Rossi}, ``On the asymptotic distribution of
  time-reversal {MUSIC} null spectrum,'' \emph{Elsevier Digital Signal
  Processing}, submitted, 2016.

\bibitem{Asgedom2011}
E.~G. Asgedom, L.-J. Gelius, A.~Austeng, S.~Holm, and M.~Tygel, ``Time-reversal
  multiple signal classification in case of noise: A phase-coherent approach,''
  \emph{The Journal of the Acoustical Society of America}, vol. 130, no.~4, pp.
  2024--2034, 2011.

\bibitem{Yavuz2008}
M.~E. Yavuz and F.~L. Teixeira, ``On the sensitivity of time-reversal imaging
  techniques to model perturbations,'' \emph{IEEE Transactions on Antennas and
  Propagation}, vol.~56, no.~3, pp. 834--843, Mar. 2008.

\bibitem{Fannjiang2011}
A.~C. Fannjiang, ``The {MUSIC} algorithm for sparse objects: a compressed
  sensing analysis,'' \emph{Inverse Problems}, vol.~27, no.~3, p. 035013, 2011.

\bibitem{Krim1996}
H.~Krim and M.~Viberg, ``Two decades of array signal processing research: the
  parametric approach,'' \emph{{IEEE} Signal Process. Mag.}, vol.~13, no.~4,
  pp. 67--94, 1996.

\bibitem{Shi2005}
G.~Shi and A.~Nehorai, ``Maximum likelihood estimation of point scatterers for
  computational time-reversal imaging,'' \emph{Communications in Information \&
  Systems}, vol.~5, no.~2, pp. 227--256, 2005.

\bibitem{Stewart1973}
G.~W. Stewart, ``Error and perturbation bounds for subspaces associated with
  certain eigenvalue problems,'' \emph{SIAM review}, vol.~15, no.~4, pp.
  727--764, 1973.

\bibitem{Liu2008}
J.~Liu, X.~Liu, and X.~Ma, ``First-order perturbation analysis of singular
  vectors in singular value decomposition,'' \emph{{IEEE} Trans. Signal
  Process.}, vol.~56, no.~7, pp. 3044--3049, Jul. 2008.

\bibitem{Bernstein2009}
D.~S. Bernstein, \emph{Matrix mathematics: theory, facts, and formulas}.\hskip
  1em plus 0.5em minus 0.4em\relax Princeton University Press, 2009.

\bibitem{Schreier2010}
P.~J. Schreier and L.~L. Scharf, \emph{Statistical Signal Processing of
  Complex-Valued Data: The Theory of Improper and Noncircular Signal}.\hskip
  1em plus 0.5em minus 0.4em\relax Cambridge, 2010.

\bibitem{Agarwal2008}
K.~Agarwal and X.~Chen, ``Applicability of {MUSIC-type} imaging in
  two-dimensional electromagnetic inverse problems,'' \emph{{IEEE} Trans.
  Antennas Propag.}, vol.~56, no.~10, pp. 3217--3223, 2008.

\bibitem{Solimene2014}
R.~Solimene and A.~Dell'Aversano, ``Some remarks on time-reversal {MUSIC} for
  two-dimensional thin {PEC} scatterers,'' \emph{IEEE Geoscience and Remote
  Sensing Letters}, vol.~11, no.~6, pp. 1163--1167, Jun. 2014.

\bibitem{Fouda2014}
A.~E. Fouda and F.~L. Teixeira, ``Statistical stability of ultrawideband
  time-reversal imaging in random media,'' \emph{IEEE Transactions on
  Geoscience and Remote Sensing}, vol.~52, no.~2, pp. 870--879, Feb. 2014.

\end{thebibliography}

\end{document}